\documentclass{article}

\usepackage[square,numbers]{natbib}
\usepackage[preprint]{neurips_2024}

\usepackage[utf8]{inputenc} 
\usepackage[T1]{fontenc}    
\usepackage{hyperref}       
\usepackage{url}            
\usepackage{booktabs}       
\usepackage{amsfonts}       
\usepackage{nicefrac}       
\usepackage{microtype}      
\usepackage{xcolor}         
\usepackage{natbib}         
\usepackage{graphicx}       
\usepackage{todonotes}
\usepackage{hyperref}       

\title{Lumename: Wearable Device for Hearing Impaired w/ Personalized
ML-Based Auditory Detection and Haptic-Visual Alerts}

\author{%
  Jeanelle Dao\textsuperscript{1} and Jadelynn Dao\textsuperscript{2,*}\\
  \textsuperscript{1}Presentation High School\\
  \textsuperscript{2}California Institute of Technology\\
  \textsuperscript{*}Corresponding author: Jadelynn Dao, jdao@caltech.edu\\
}

\begin{document}

\maketitle

\begin{abstract}
  According to the World Health Organization, 430 million people experience disabling hearing loss. For them, recognizing spoken commands such as one's name is difficult. To address this issue, Lumename, a real-time smartwatch, utilizes on-device machine learning to detect a user-customized name before generating a haptic-visual alert. During training, to overcome the need for large datasets, Lumename uses novel audio modulation techniques to augment samples from one user and generate additional samples to represent diverse genders and ages. Constrained random iterations were used to find optimal  parameters within the model architecture. This approach resulted in a low-resource and low-power TinyML model that could quickly infer various keyword samples while remaining 91.67\% accurate on a custom-built smartwatch based on an Arduino Nano 33 BLE Sense. 
  
\end{abstract}

\section{Introduction}

\subsection{Background and Significance}

430 million people, including 34 million children, have disabling hearing loss \cite{who2021hearing}. This makes everyday communication and interaction particularly difficult. This is increasingly difficult when those with hearing loss cannot see who are talking to them, particularly when being called. The lack of visual cues worsens the communication barrier, further impacting everyday activities.

\subsection{Objectives}

The objective of this project is to assist individuals with hearing loss in recognizing spoken commands by utilizing on-device machine learning to detect a user's customized keyword  via a real-time smartwatch.  Subsequently, a haptic and visual alert will be issued to alert the user.

The custom keyword will be chosen by the user, and the user will be prompted to upload recordings of their chosen keyword (ex. their specific name) to the cloud for training. Therefore, the framework to build the model must be general enough to work for any given word, yet the final model should be specific enough to avoid false positives.

To be effective, this device must be at least 85\% accurate when testing how well it identifies the correct keyword (true positives) and ignores unrelated sounds (true negatives). Computing resources must be consciously limited to ensure the device can run for a minimum of eight hours and respond to the keyword within three seconds without internet access.

\section{Methods}

\subsection{Data Preprocessing}

Samples are 16-bit at 1kHz audio data that are organized into eight total subclasses. The two parent classes are “name” and “not name,” and each class contains four subclasses. “Name” is split based on the method of audio transformation: \textit{pitch shift}, \textit{ambiance mixing}, \textit{both}, or \textit{neither}. However, other transformations such as speed, volume, and timing (which shifts the audio to play earlier or later than the original) are present in every “name” sample.

Conversely, “not name” is split into different types of background noise: \textit{static}, \textit{ambiance}, \textit{words}, and \textit{words with ambiance mixing}. The \textit{words} subclass is made of random samples from the Google Speech Commands dataset \cite{speechcommandsv2}. Each of these \textit{word} samples are transformed similarly to the “name” samples. The \textit{static} samples are generated from the microphone used in the final device, and serve as a baseline recording of silence.

\begin{figure}[htbp]
    \centering
    \fbox{\includegraphics[width=\textwidth,height=\textheight,keepaspectratio]{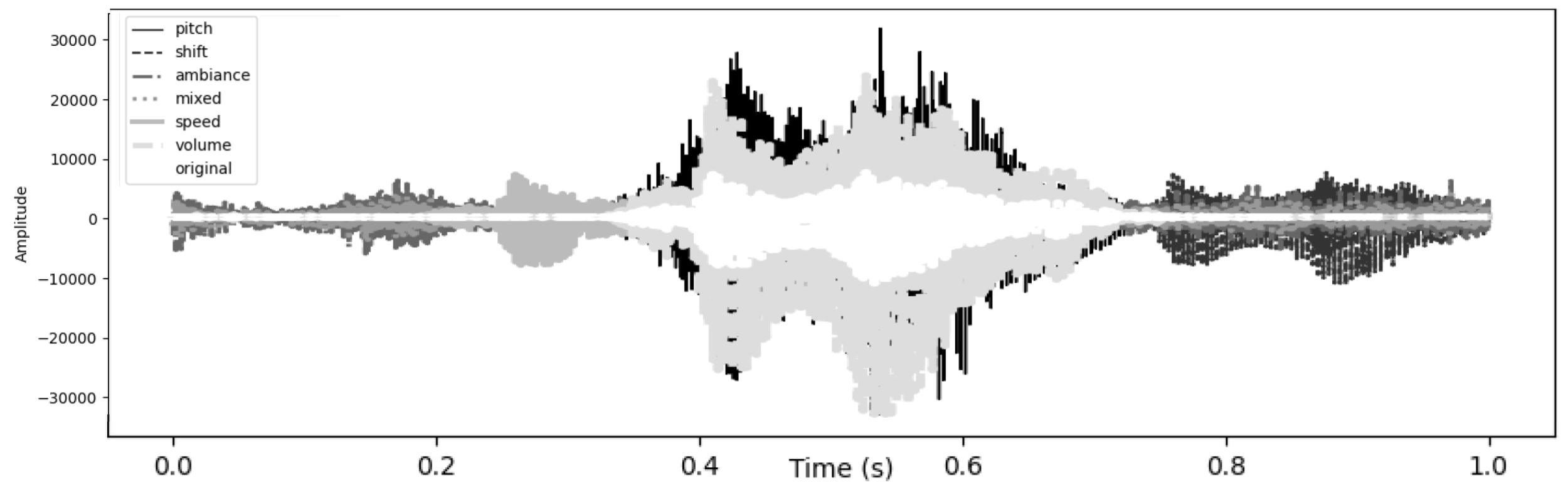}}
    \caption{sample waveform shifted by audio manipulation code}
    \label{fig:1}
\end{figure}

Originally, the "name" dataset only consists of the limited samples recorded by the user.  While audio augmentation is common practice to expand datasets \cite{ko15_interspeech}, it is not commonly used to represent completely different voices. However, for Lumename, an audio manipulation code transforms the user’s samples into a diverse dataset emulating additional genders and ages. The program randomly shifts a sample’s pitch, speed, volume, and timing (within a one second segment). Additionally, royalty-free ambiance sounds \cite{pixabay}, which are permitted for use under Pixabay's Content License, are mixed into  samples to train for different sets of background noise (see Figure \ref{fig:1}).

Additionally, samples are uploaded to the cloud into the training framework, Edge Impulse \cite{hymel2023edge}. Using the Edge Impulse API, samples are then preprocessed using mel-frequency cepstral coefficients (MFCC), which transforms data into a representation that highlights important features and removes insignificant data \cite{MFCC}. Utilizing a constrained random .json file, twelve different combinations of MFCC coefficient parameters were tested, with a focus on frame length, frame stride, and number of filters. The final parameters (0.032s, 0.032s, and 40 filters respectively) contained the best balance between containing a sufficient amount of data, preventing overfitting, and reducing necessary compute power.

\subsection{Model Training}

As the model architecture would be run directly on a limited-resource microcontroller, the model needed to be created with maximal simplicity while remaining accurate, which was mainly achieved by adjusting the number of layers used \cite{hiddenlayers}. Therefore, by leveraging a constrained random .json file, over 72 combinations of low-resource layers were trained using Edge Impulse. After testing both 1D and 2D convolution, 1D convolution ultimately performed well and required less computing resources. Within both convolution types, different amounts of filters were also tested. Dropout layers were also utilized to prevent overfitting, with the dropout rate being adjusted within the constrained random loop.

To determine which sets of parameters performed better, a validation set was constructed using non-user generated samples. This ensured that the final model was general enough to detect when anyone said the keyword, not just the user. Additional static, ambiance, and Google Speech Command samples (that were not in the training set) were also included in the aforementioned validation set.

\begin{figure}[htbp]
    \centering
    \fbox{\includegraphics[width=\textwidth,height=\textheight,keepaspectratio]{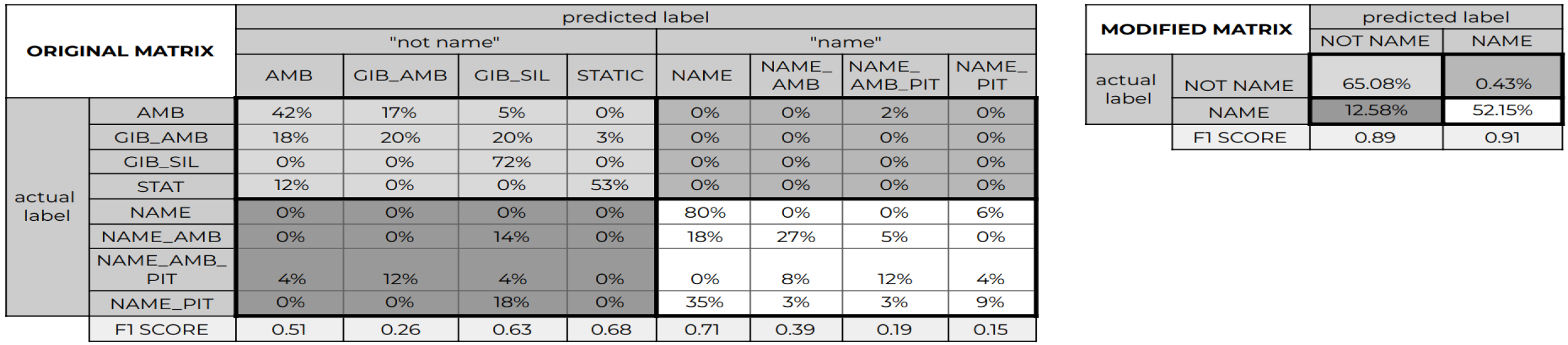}}
    \caption{example of original and modified confusion matrix}
    \label{fig:2}
\end{figure}

To determine performance, a standard confusion matrix is generated using the subclass labels in the original training and validation sets. Subsequently, subclasses were grouped under their parent class. Therefore, while accuracy between subclasses may be low, as long as samples are in the same parent class, the sample will be correctly identified. For example, if \textit{static} is classified as an \textit{ambiance}, the sample is still correctly classified as it is ultimately labeled as “not name.” Upon generating this modified confusion matrix based on parent classes (Figure \ref{fig:2}), the model with the highest F1 score is ultimately chosen as the highest-performing model. 

The final model begins with a reshape layer that transforms data processed by MFCC. It also includes two 1D convolution layers with 32 filters and 3 kernels. Each of these convolution layers is followed by a max-pooling layer to reduce dimensionality and a dropout layer to prevent overfitting \cite{Bez2019}. This final model, deployed as a TinyML model, ensures an efficient solution for real-time keyword detection on a low-power device \cite{warden2020tinyml}.

\subsection{Model Inference}

Lumename utilizes sliding-window aggregation, allowing recent and past data to be properly reflected \cite{Tangwongsan2020}. Data is streamed from the microphone in 250ms packets. At inference time, four of these packets are combined to create one second samples. These one second samples are then classified using the previously created TinyML model, where every subclass contains a corresponding probability value from 0-100\%. If the combined probabilities of the “name” subclasses are greater than an empirically predetermined threshold (\(\approx70\%\)), a light and vibration motor are triggered.

\section{Experiments}

\subsection{Variable Groups}

For positive testing, which tests how well the model can correctly classify keyword samples, the model is tested against auxiliary  samples of real-life people of different demographics. These samples represent varying pitches and background noise environments. These act as the variable groups.

For negative testing, which tests how well the model can ignore unrelated sounds, the three groups are: words from the Google Speech Commands dataset, static, and ambiance. To ensure a comprehensive evaluation of the model, these samples were not included in the training nor validation sets.

Each of these samples used in the test are prerecorded and included in a comprehensive audio recording, ensuring a replicable environment. When conducting the test, a computer will play the comprehensive recording to trigger (or to not trigger) the device.

\subsection{Main Experiments}

The purpose of these experiments are to test the device’s accuracy, response time, and  battery life. Accuracy tests are split into two categories: positive and negative. 

In positive testing, over fifty samples and five trials demonstrate that the device correctly classifies the keyword 90.00\% of the time. The greater part of incorrectly classified positive samples were ones with pitch modulation. Negative testing involves over 150 one-second samples and five trials. After testing, the device incorrectly classified negative samples an average of 6.67\% of the time.

The response time test involved timing the length between the spoken keyword and the device’s haptic-visual alert. Measuring down to the millisecond, ten trials reveals the response time being an average of 0.906 seconds.

\begin{figure}[htbp]
    \centering
    \fbox{\includegraphics[width=\textwidth,height=\textheight,keepaspectratio]{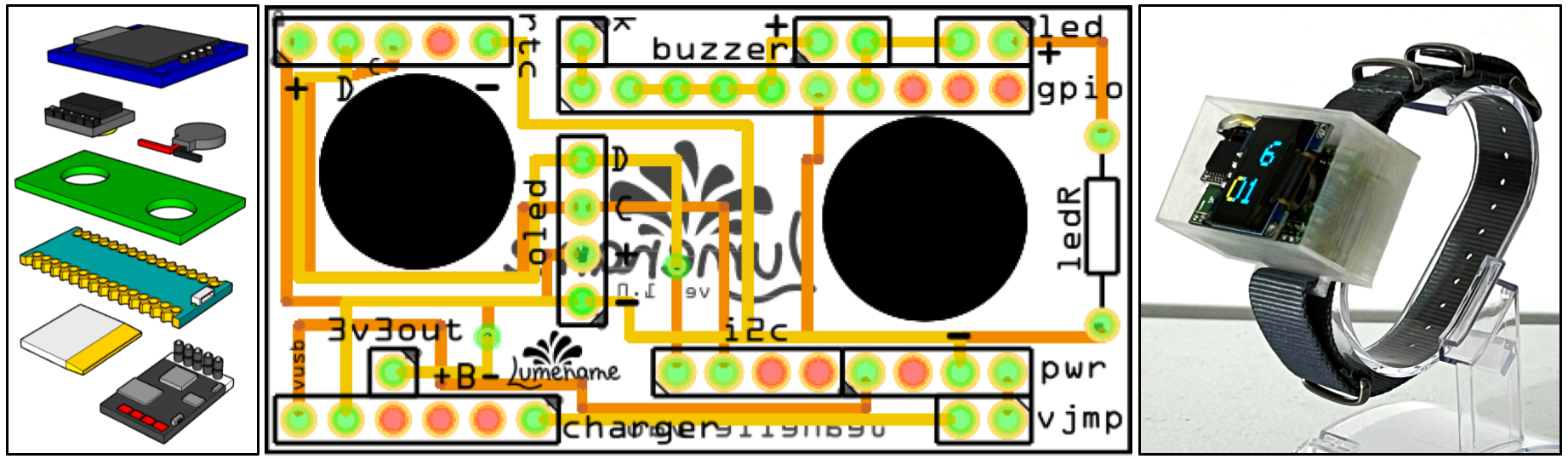}}
    \caption{(left to right) diagram of components, custom PCB layout, assembled prototype}
    \label{fig:3}
\end{figure}

To measure the device’s battery consumption, an Arduino Nano 33 BLE Sense, a 150mAh LiPo battery, and other miscellaneous parts, were employed to construct a custom smartwatch prototype used for reference (see Figure \ref{fig:3}). This test estimated a daily usage of 100 buzzes to provide a wide margin. Measurements indicated the device can last beyond eight hours, even with the high margins. Followed by a real-world test, the device was placed in a silent room to increase replicability. Without buzzes, the device lasts beyond eleven hours. Because one hundred buzzes only consume 0.4\% of the battery, it was concluded that this could last a normal day of wear for an average user.

\section{Discussion}

\subsection{Implication of Results}

Lumename suggests the possibility of using on-device machine learning for custom keyword recognition, even on limited-resource microcontrollers. By demonstrating the effectiveness of using audio manipulation to mimic other voices, the specified method of expanding small datasets provides a viable solution for creating personalized models using limited data. 

Moreover, security concerns about continuous microphone recording \cite{EECSprivacy} for a keyword are reduced in this project. While running on-device, Lumename does not save any data after the sliding window has passed a sample, as data is overwritten. In addition, because the wearable device runs separately from the cloud, there is no risk of cloud data breaches. 

\subsection{Limitations and Challenges}

While the data was manipulated in pitch, speed, and timing to mimic other human voices, other vocal qualities that drastically affect word perception, such as nasality and breathiness \cite{vocalquality}, may not be sufficiently represented in the dataset. Therefore, the model may still face challenges in accurately recognizing the keyword across different speakers and conditions. Additionally, while the model has gone through testing, it was evaluated using a limited dataset. The model still requires more extensive testing, validation in real-world settings, and with a wider variety of custom keywords.

\subsection{Future Work Recommendations}

A major step towards improving this project is increased vocal mimicking. As noted in Section 4.2, representation of important vocal qualities are still missing. As keywords are meant to be user-customized, a variety of keyword options and languages require testing. Examining performance of such a variety will also provide valuable information about any spurious correlation affecting the model's performance with certain sounds. Consequently, approaches such as feature disentanglement or additional audio manipulation may need to be employed to reduce correlation \cite{ye2024spurious}.

\bibliography{references}

\end{document}